\documentclass[baaa]{baaa}

\usepackage[pdftex]{hyperref}
\usepackage{subfigure}
\usepackage{natbib}
\usepackage{helvet,soul}
\usepackage[font=small]{caption}

\contriblanguage{1}

\contribtype{1}

\thematicarea{5}

\title{Ionizing photons produced by massive stars in SMC-N88a}

\titlerunning{Ionizing photons estimation in massive stars}

\author{
M.T. Krilich\inst{1}\&
C.G. D\'iaz\inst{2,3}
}

\authorrunning{Krilich et al.}

\contact{tomas.krilich@mi.unc.edu.ar}

\institute{
Facultad de Matemática, Astronomía, Física y Computación, UNC, Argentina
\and
Observatorio Astron\'omico de C\'ordoba, UNC, Argentina
\and
Consejo Nacional de Investigaciones Cient\'ificas y T\'ecnicas, Argentina
}

\resumen{ La regi\'on H {\sc ii} N88a en la Nube Menor de Magallanes (NmM) es una regi\'on esf\'erica de $1.5~\mathrm{pc}$ de diámetro, con alta densidad de gas y polvo, y al menos cuatro
estrellas masivas en su interior. 
Estudios previos sugieren que las fuentes conocidas podrían
ser insuficientes para ionizar la regi\'on y explicar la emisi\'on nebular. 
En esta contribuci\'on analizamos la producci\'on de fotones ionizantes de las cuatro fuentes conocidas dentro de la región H {\sc ii}.
Comparamos la fotometría disponible en la bibliograf\'ia con la distribuci\'on espectral de energ\'ia de los modelos “Potsdam Wolf-Rayet” (PoWR) de estrellas masivas
para un amplio rango de coeficientes de extinci\'on por polvo (Av).
En particular, seleccionamos modelos
de atm\'osferas de tipo OB con la metalicidad de la NmM, y comparamos la predicci\'on  del flujo
de fotones ionizantes con estimaciones basadas en la emisi\'on nebular de la región H {\sc ii}.
 Encontramos que los valores de Av que mejor reproducen la fotometr\'ia de cada fuente var\'ian entre 0.82 y 3.84, aumentando hacia la banda de polvo que atravieza a la nebulosa. Adem\'as, las fuentes son compatibles con estrellas tipo O con $\mathrm{T_{eff}} > 40 ~\mathrm{kK}$ y cocientes intr\'insecos $\mathrm{F(LyC)/F(UV)}>1$. 
 Finalmente, la tasa total de fotones ionizantes predicha por los modelos m\'as calientes es $\mathrm{log(Q_H)} > 49.6 ~\mathrm{ph\ s}^{-1}$, lo que sugiere que las estrellas podr\'ian mantener el estado de ionizaci\'on de la nebulosa.}

\abstract{The H {\sc ii} region N88a in the Small Magellanic Cloud (SMC) is a spherical region of  $1.5~\mathrm{pc}$ diameter, with high concentration of gas and dust, and at least four massive stars within it. Previous studies suggest that the four known sources may be
insufficient to ionize the region and explain the nebular emission. 
In this contribution we analyze the ionizing photon production of the four known sources within the H {\sc ii} region. We compared the available photometry in the literature with the spectral energy distribution calculated for the ``Potsdam Wolf-Rayet'' (PoWR) models of massive stars
for a wide range of dust extinction coefficient (Av).
In particular, we selected models of OB-type atmospheres with SMC metallicity and compared the ionizing photon flux prediction with previous estimates based on the nebular emission of the H {\sc ii} region. 
We found that the Av values that best reproduce the photometry of each source vary from 0.82 to 3.84, increasing toward the dust band that runs through the nebula. 
In addition, all four sources are compatible with O-type stars with $\mathrm{T_{eff}} > 40 ~\mathrm{kK}$ and intrinsic ratios $\mathrm{F(LyC)/F(UV)}>1$. 
Lastly, the total ionizing photon rate predicted by the hottest models is $\mathrm{log(Q_H)} > 49.6  ~\mathrm{ph\ s}^{-1}$, which suggests that the stars could maintain the ionization state of the nebula.
}


\begin{document}

\maketitle

\section{Introduction}\label{S_intro}

According to the dominant cosmological model, intergalactic hydrogen became fully ionized one billion years after the Big Bang (redshift $z\approx6$), during the ``epoch of reionization''. 
Massive stars in low-mass galaxies are suspected to be the main sources of ionizing photons 
\citep[e.g.,][]{Robertson2013,henkel2022}. 
However, the escape fraction ($f_{esc}$) of hydrogen ionizing photons (Lyman continuum, LyC $\lambda< 912$~\AA) from the sources to the intergalactic medium is unknown. This radiation is not observed in objects at $z > 5$ due to absorption by neutral hydrogen. Thus, several $f_{esc}$ indicators have been proposed at different wavelengths, from the Lyman-$\alpha$ emission line profile at $\lambda = 1216$~\AA\ \citep{Verhamme2015,Gazagnes2020} to infrared lines at $\lambda = 120$~$\mu $m \citep{Ramambason2022}. 
These indicators are based on the ``relative'' escape fraction \citep{Steidel2001,Steidel2018}, calculated as:

\begin{equation}
f_{esc,rel} = \frac{\langle
\mathrm{F_{LyC}/F_{UV}}\rangle_{out}}{\langle \mathrm{F_{LyC}/F_{UV}}\rangle_{in}}10^{[A_{\lambda}(900)-A_{\lambda}(1500)]/2.5},
\end{equation}

\vspace{0.2 cm}
\noindent where the observed output ($out$) and the unobserved intrinsic ($in$) flux ratios are compared, for $\lambda_{rest}= 900$~\AA\ (LyC) and $\lambda_{rest}= 1500$~\AA\ (UV). 
Therefore, the interpretation of $f_{esc,rel}$ from galaxy observations depends on the intrinsic (unobserved) average ratio $\langle F_{\mathrm{LyC}}/F_{\mathrm{UV}}\rangle_{in}$ assumed for the observed source. 

Compact {H {\sc ii} regions in low-mass low-metallicity galaxies are
excellent laboratories to explore the escape of
ionizing radiation from young massive stars
in conditions that are commonly assumed for galaxies 
driving the epoch of reionization.
We selected SMC-N88a in the Small Magellanic Cloud to study 
the predictive power of $f_{esc}$ indicators based on nebular emission lines \citep{diaz2023}. The main goal is to improve
the interpretation of $f_{esc,rel}$ from the nebular emission observed in galaxies in the epoch of reionization. 

In this contribution, we report the intrinsic flux ratio 
$\langle \mathrm{F_{LyC}/F_{UV}}\rangle_{in}$ predicted for the four stars likely inside SMC-N88a \citep{testor2010}. For this purpose, we identified OB stellar atmosphere models and Av extinction values that best reproduce the available photometry. Finally, we report the ionizing photon production rate and the ionizing LyC and non-ionizing UV fluxes, predicted by the top ranked models for each star.

\section{SMC-N88a}

The object SMC-N88a is a compact high-excitation  H {\sc ii} region, also known as ``High Excitation Blob'', which has a circular shape with a diameter of approximately $1.5~\mathrm{pc}$, located in the Magellanic bridge near the SMC.
According to \citet[][HM99 hereafter]{hm1999_n88},
the nebular emission
H$\beta = 1.85\times 10^{-11}~\mathrm{erg\ s}^{-1} \mathrm{cm}^{-2}$ corresponds to $\log(\mathrm{Q_H})= 49.32 ~\mathrm{ph\ s}^{-1}$.
HM99 suggests that a single O6V star is sufficient to ionize this region. However, \citet{Indebetouw2004}, studying the radio emission at $3 ~\mathrm{cm}$, argued that the region would need an O4V-O5V type and $\log(\mathrm{Q_H})= 49.5 ~\mathrm{ph\ s}^{-1}$.
\citet{testor2010} report J, H, and Ks magnitudes of four sources potentially inside the region, whose colors in the observed bands are consistent with massive main sequence stars. They also analyzed Ks band spectra and did not detect the He {\sc ii} $2.185 ~\mathrm{ \mu m}$ absorption line nor the N {\sc iii} $2.115~\mathrm{ \mu m}$ line in any of the sources. This suggests that the stars would be cooler than O8V and at least five of such stars would be needed to produce the minimum flux of ionizing photons required. Furthermore, in a case of $f_{esc} > 0$, it is reasonable that the rate of ionizing photon production is higher than the estimated from the reprocessed nebular emission, since it must take into account the emission that escapes from the region and is not absorbed by the nebula. The discrepancy between the nebular emission and the stars observed inside the nebula suggests that the ionizing sources are insufficient. Some of the reasons could be: unidentified sources inside the nebula, ionizing sources external to the nebula, or that the temperature estimates have a large uncertainty, partly due to dust extinction.

\section{Analysis}\label{sec:guia}

In this work, we present a parameter-space analysis of the stellar atmosphere models ``Potsdam Wolf-Rayet'' \citep[PoWR,][]{PoWR} 
that best reproduce the photometry of the four stars within the target \citep[namely stars 37, 41, 42, and 47 from][]{testor2010}.
The main questions are the following:
\begin{itemize}
 \item Is the photometry reported by \citet{testor2010} compatible with ionizing sources? What is the effective temperature range?
 \item What is the dust extinction that best reproduces the observations?
 \item What is the ionizing photon production rate predicted by current models for the available observations of the four sources?
 \item What is the intrinsic ratio  
$\mathrm{F_{LyC}/F_{UV}}$ predicted for each case?
\end{itemize}

\begin{figure}
    \centering
    \includegraphics[width=7cm, height=8cm]{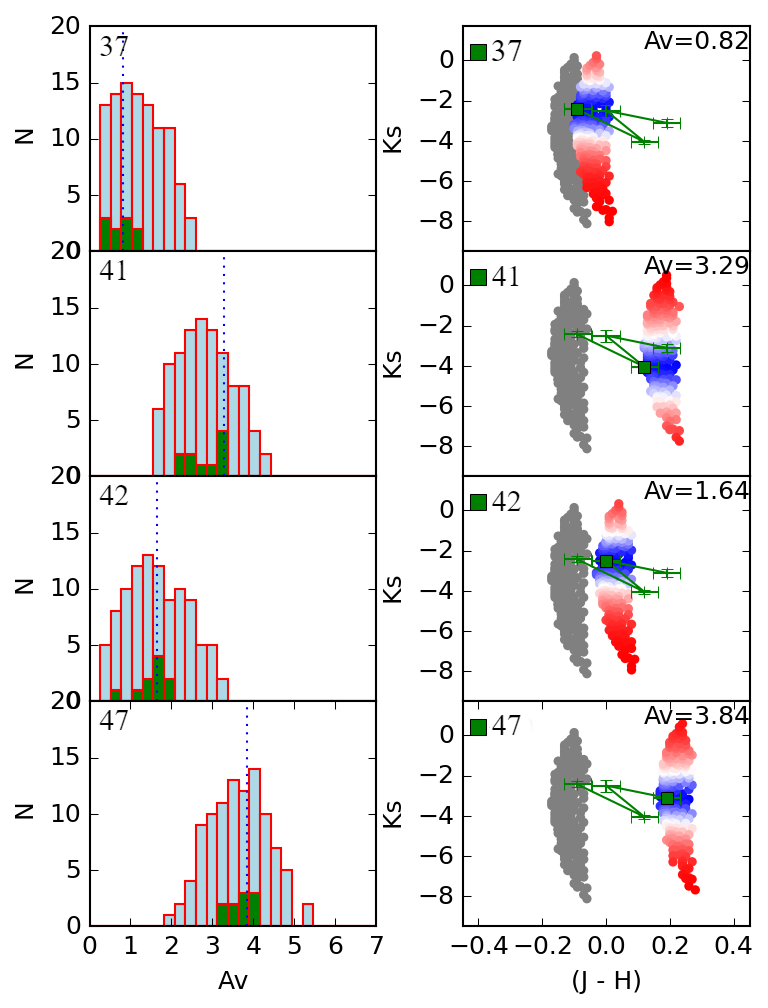}
    \caption{ \textit{Left}: Av parameter distribution of the best models for each star: top 10 green histograms, top 100 blue histograms.
    \textit{Right} : Color-magnitude diagram of the sample stars (green), models without extinction (gray), and with extinction (color). The color indicates the order of merit from blue to red.}
    \label{histogramas}
\end{figure}

Firstly, we obtained the spectral energy distribution (SED) of OB-type stars, calculated with SMC metallicity and variable mass-loss rate according to \citet{vink}, from the official site \url{www.astro.physik.uni-potsdam.de/~wrh/PoWR/SMC-OB-Vd3}.
Secondly, we applied dust extinction to each SED according to the SMC average extinction curve of \citet{gordon2003}
using the \textsc{dust${-}$extinction} (v1.2) package of \textsc{astropy} \citep{astropy}.
Assuming $\mathrm{Rv} = 2.74$  \citep[the average Rv for the SMC bar from ][]{gordon2003}, we tested Av values from 0.27 to 13.43 (i.e. $\mathrm{E(B-V)} = 0.1$ to 4.9).
Thirdly, for each model with the applied extinction, we calculated the absolute magnitudes in the J, H, and Ks bands of the 2MASS system  with the \textsc{pyphot} package \citep{fouesneau2022}, 
assuming a distance of $66 ~\mathrm{k pc}$.
Finally, the models for each star
were ranked according to the residuals: $\mathrm{S = ((J-H)_M-(J-H)_O)^2+(Ks_M-Ks_{O})^2}$, which compares the observed color $\mathrm{(J-H)_O}$ \citep[Table 3 of ][]{testor2010} with the model
color $\mathrm{(J-H)_M}$, and the absolute Ks magnitude corrected for nebular emission \citep[Ks$_{PSF}$, Table 4 of ][]{testor2010}
with the absolute Ks magnitude of the model.


From this ranking, we selected Av values that represent the top 10 models.  Figure \ref{histogramas} displays the Av distribution of the top 100 models (blue bars) and the top 10 models (green bars) for each source.
Except for star 41, the peak of the top 10 Av distributions (vertical dotted line in Fig. \ref{histogramas}) agrees with the peak of the top 100. Therefore, we selected the models at the peak of the Av distribution for each star.
These models are presented in Fig. \ref{param} with colored dots indicating their order of merit from blue to red.

\begin{table*}[!t]
\centering
\caption{ Best ranked models for the four massive stars within SMC-N88a.
Column 1: Star ID, Column 2:  selected Av, Column 3: Model ID, Column 4: ionization rate, Columns 5 and 6: LyC and UV fluxes, and Column 7: intrinsic flux ratio. }
    \begin{tabular}{lccccccc}
    \hline\hline\noalign{\smallskip}
    Star & Av &Model& $\log(\mathrm{Q_H})$& $\mathrm{F_{LyC}}$ &$\mathrm{F_{UV}}$& $\mathrm{F_{LyC}/F_{UV}}$\\
    &  \!\![mag]& \!\!\! & \!\! [ph \,$\mathrm{s}^{-1}$]&[$10^{-6}\mathrm{erg}$\,\AA$^{-1}\mathrm{cm}^{-2}\mathrm{s}^{-1}$]&[$10^{-6}\mathrm{erg}$\,\AA$^{-1}\mathrm{cm}^{-2}\mathrm{s}^{-1}$]& \\
    \hline\noalign{\smallskip}
    \!\!37 & 0.82  & 40-44      & 48.34&    9.55$\pm$0.03   &6.33$\pm$0.02&1.5\\
                   &&30-40      & 47.22&    1.370$\pm$0.005  &4.43$\pm$0.08&0.3\\
                   &&25-38      & 47.91&   0.0280$\pm$0.002  &2.72$\pm$0.03&0.1\\
    \hline
    \!\!41 & 3.29   & 44-40     & 49.38&    84.8$\pm$0.3  &34.1$\pm$0.2&2.6\\
                    &&38-38     & 49.07&    55.6$\pm$0.1  &28.4$\pm$0.6& 2.0\\
                    &&33-36     & 48.66&   30.99$\pm$0.02 &23.9$\pm$0.8&1.3\\
                    &&28-34     & 47.91&    7.08$\pm$0.02 &19.3$\pm$0.4&0.4\\

    \hline
    \!\!42 & 1.64   & 42-44     & 48.54&    13.44$\pm$0.05 &7.657$\pm$0.008&1.8\\
                    &&36-42     & 48.11&     7.49$\pm$0.01 &6.31$\pm$0.07&1.2\\
                    &&31-40     & 47.46&    2.295$\pm$0.005&5.1$\pm$0.1&0.4\\
                    &&26-38     & 46.67&    0.460$\pm$0.003&3.48$\pm$0.03&0.1\\
    \hline
    \!\!47 & 3.84   & 43-42     & 48.95&    32.4$\pm$0.1    &16.03$\pm$0.01&2.0\\
                    &&37-40     & 48.60&   20.91$\pm$0.04   &13.6$\pm$0.2&1.6\\
                    &&32-38     & 48.08&    8.830$\pm$0.005  &11.5$\pm$0.3&0.8\\
                    
    \hline
    \end{tabular}
\label{tabla1}
\end{table*}

Finally, the average LyC flux in the range $812-912$~\AA\ and the UV flux between $1450-1550$~\AA\ were calculated with \textsc{pyphot} for each star (Fig. \ref{lyc_uv}), and the intrinsic ratio $\mathrm{F_{LyC}/F_{UV}}$ was obtained. The values are presented in Table \ref{tabla1}. Since the template spectra have no reported uncertainties,
the errors 
represent the uncertainties due to data binning and edge limits during the integration process.
Ionization rates were extracted from the official website of the models, which does not report the uncertainties.

\begin{figure*}[!t]
    \centering
    \includegraphics[scale=0.5]{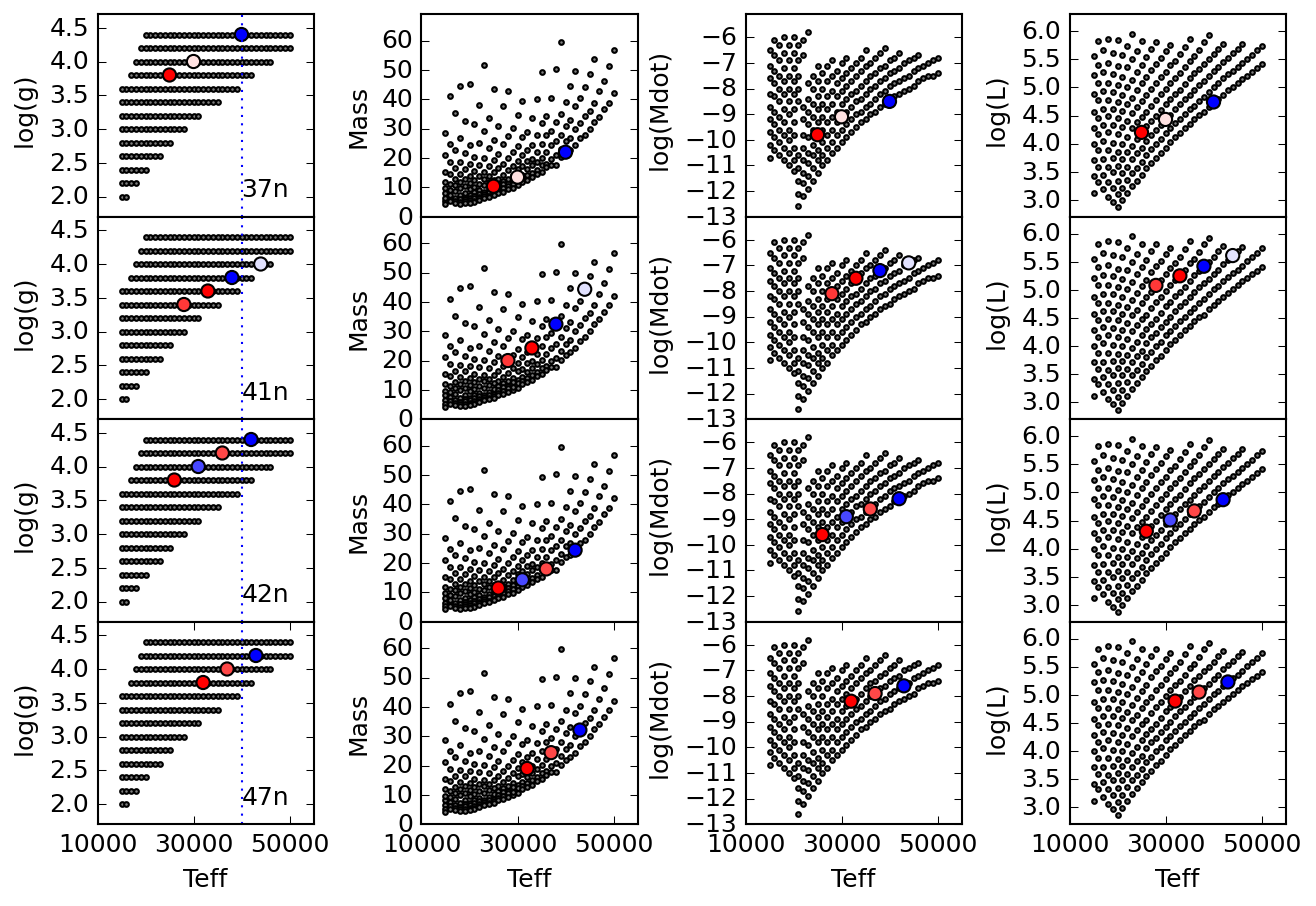}
    \caption{Parameter-space of the PoWR models used in the analysis (gray dots). Each row corresponds to a different star.  The models in the top 10, which also exhibit the average extinction (as indicated in Fig. \ref{histogramas}), are color-coded based on the order of merit, with blue representing the first and red the last. All four stars are compatible with  $\mathrm{T_{eff}}>40 ~\mathrm{kK}$, which are the preferred models (blue dots) for stars 37, 42, and 47.}
    \label{param}
\end{figure*}

\begin{figure}
    \centering
    \includegraphics[width=8cm, height=5cm]{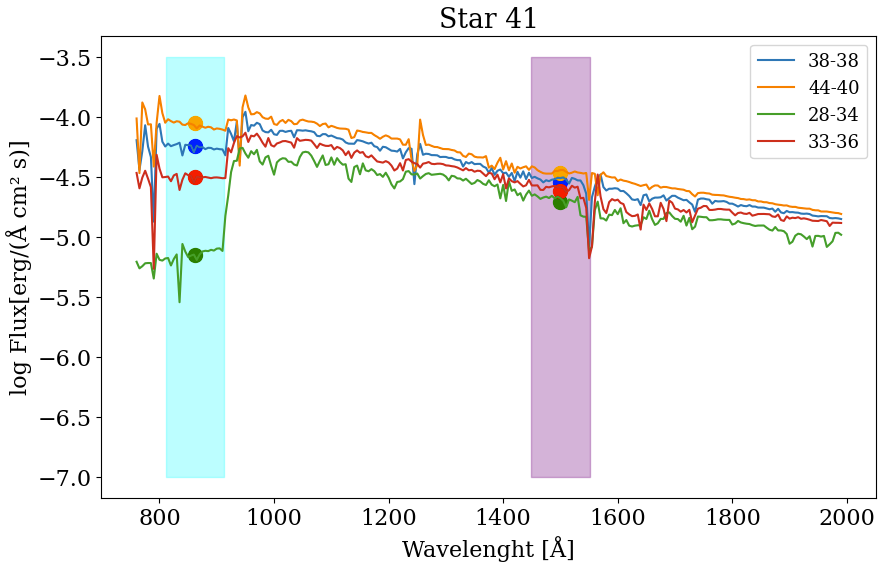}
    \caption{UV spectra of the top-ranked models for star 41. The vertical color bands indicate the spectral regions where the fluxes reported in Table \ref{tabla1} were measured. LyC flux was measured between $812 - 912$~\AA\ and UV continuum flux between $1450 - 1550$~\AA\ .}
    \label{lyc_uv}
\end{figure}

\section{Results}
 From our analysis we report the following results.
\begin{itemize}
 \item Dust extinction: The Av values that best reproduce the photometry of each star vary from 0.82 to 3.84 (Fig. \ref{histogramas}), increasing from southwest to northeast, i.e.  toward the dust band seeing across the region.
 \item Comparison with PoWR models: Assuming the values of Av from Table \ref{tabla1}, most of the models that best reproduce the photometric conditions have $\mathrm{T_{eff}} > 30 ~\mathrm{kK}$, corresponding to spectral type O. In particular, the best ranked model for stars 37, 42, and 47 have $\mathrm{T_{eff}} = 40 ~\mathrm{kK}$, $42 ~\mathrm{kK}$, and $43 ~\mathrm{kK}$, respectively.

  For star 41, the second best model has $\mathrm{Teff}=44 ~\mathrm{kK}$, and it is the most ionizing model with $\mathrm{log(Q_H)} = 49.38 ~\mathrm{ ph\ s}^{-1}$.
 \item Ionizing photon rate:  Two of the top models have $\mathrm{log(Q_H)} > 49  ~\mathrm{ph\ s}^{-1}$ and both correspond to star 41. Moreover, the total ionizing photon rate
 summed over the hottest models of each star is $\mathrm{log(Q_H)} = 49.6~\mathrm{ph\ s}^{-1}$.
 \item LyC and UV fluxes:  The four stars are compatible with ratios $\mathrm{F(LyC)/F(UV)}$  $>1$. 
\end{itemize}

\section{Conclusions}

In this work we provide new evidence that star 41 could be the main source of LyC photons as the best ranked models reach $\mathrm{log(Q_H)} > 49 ~\mathrm{ph\ s}^{-1}$. Moreover, all 4 stars are compatible with spectral type O, and the combined contribution from the hottest models of each star reaches $\mathrm{log(Q_H)} = 49.6 ~\mathrm{ph\ s}^{-1}$.
This would be enough to account for the $\mathrm{H\beta}$ emission observed in the nebula, as reported by \citet{hm1999_n88}, with $\mathrm{log(Q_H)}= 49.32 ~\mathrm{ph\ s}^{-1}$, and for the radio emission observed by \citet{Indebetouw2004}, which requires $\mathrm{log(Q_H)}= 49.5 ~\mathrm{ph\ s}^{-1}$.

Assuming that the ionization sources are located in the interior of the H {\sc ii} region, our analysis of the PoWR models indicates that the combined sources could be enough to explain the ionization state of the nebula. In addition, we report a possible extinction gradient, consistent with the position of the absorption band and the hypothesis that the stars are inside the region. This spatial variability in Av is indicative of a multiphase environment with a heterogeneous dust distribution on sub-parsec scales. In this framework, 
considering that solutions with larger dust extinction show a tendency for hotter models, 
our findings suggest that the apparent ionizing photon deficit is 
partially due to underestimated dust extinction.

\begin{acknowledgement}
MK acknowledges the support of the Observatorio  Astron\'omico de C\'ordoba and the Facultad de Matem\'atica, Astronom\'ia, F\'isica y Computaci\'on for providing travel funds to asisst to the 65 RAAA.  MK  also thanks  the Observatorio Astron\'omico de C\'ordoba for providing the working space for this reasearch.
 CGD  acknowledges the support of the ``Agencia Nacional de promoci\'on de la investigaci\'on, el desarrollo tecnol\'ogico y la innovaci\'on'' through project PICT-2021-GRF-TII-00442, and the support of CONICET through project PIP 11220210100520CO.
\end{acknowledgement}


\bibliographystyle{baaa}
\small
\bibliography{starsn88a}
 
\end{document}